\documentclass[11pt]{article}
\pdfoutput=1
\usepackage{jheppub}
\usepackage[T1]{fontenc}

\def\beq{\begin{equation}}
\def\eeq{\end{equation}}
\def\bec{\begin{center}}
\def\eec{\end{center}}

\newcommand{\bea}{\begin{eqnarray}}
\newcommand{\eea}{\end{eqnarray}}
\newcommand{\non}{\nonumber}
\newcommand{\mc}{\mathcal}

\newcommand{\mr}{\mathrm}

\def\muf{\mu_\mr{F}}
\def\mur{\mu_\mr{R}}

\def\({\left(} 
\def\){\right)} 

\def\neu{\tilde{\chi}}
\def\usqr{\tilde{u}_{R}}
\def\dsqr{\tilde{d}_{R}}
\def\usql{\tilde{u}_{L}}
\def\dsql{\tilde{d}_{L}}
\def\csqr{\tilde{c}_{R}}
\def\ssqr{\tilde{s}_{R}}
\def\csql{\tilde{c}_{L}}
\def\ssql{\tilde{s}_{L}}
\def\tsqr{\tilde{t}_{2}}
\def\bsqr{\tilde{b}_{2}}
\def\tsql{\tilde{t}_{1}}
\def\bsql{\tilde{b}_{1}}

\newcommand{\GeV}{{\mathrm{GeV}}}
\newcommand{\TeV}{{\mathrm{TeV}}}

\def\PROSPINO{{\tt PROSPINO}}
\newcommand{\RESUMMINO}{{\tt RESUMMINO}}
\newcommand{\amc}{{\tt MadGraph5\char`_aMC@NLO}}
\newcommand{\POWHEG}{{\tt POWHEG}}
\newcommand{\POWHEGBOX}{{\tt POWHEG-BOX}}
\newcommand{\PYTHIA}{{\tt PYTHIA}}

\newcommand{\PWGPY}{{\tt POWHEG+PYTHIA}}
\newcommand{\NLOPS}{{\tt NLO+PS}}
\newcommand{\LOPS}{{\tt LO+PS}}

\title{\bf 
Electroweakino pair production 
at the LHC: 
NLO SUSY-QCD corrections and parton-shower effects}

\author{Julien Baglio,}
\author{Barbara J\"ager}
\author{and Matthias Kesenheimer}
\affiliation{Institute for Theoretical Physics,  University of T\"ubingen, Auf der Morgenstelle 14, 72076~T\"ubingen, Germany}
\emailAdd{julien.baglio@uni-tuebingen.de}
\emailAdd{barbara.jaeger@uni-tuebingen.de}
\emailAdd{matthias.kesenheimer@uni-tuebingen.de}

\abstract{We present a set of NLO SUSY-QCD calculations for the pair production of neutralinos and charginos at the LHC, and their matching to parton-shower programs in the framework of the \POWHEGBOX{} program package. The code we have developed provides a SUSY Les Houches Accord interface for setting supersymmetric input parameters. Decays of the neutralinos and charginos and parton-shower effects can be simulated with \PYTHIA. To illustrate the capabilities of our program, we present phenomenological results for a representative SUSY parameter point. We find that NLO-QCD corrections increase the production rates for neutralinos and charginos significantly. The impact of parton-shower effects on distributions of the weakinos is small, but non-negligible for jet distributions.}

\arxivnumber{1605.06509}

\begin{document}

\thispagestyle{empty}
\def\thefootnote{\fnsymbol{footnote}}
\setcounter{footnote}{1}

\setcounter{page}{0}
\maketitle
\flushbottom

\def\thefootnote{\arabic{footnote}}
\setcounter{footnote}{0}
%
%%%%%%%%%%%%%%%%%%%%%%%%%%%%%%%%%%%%%%%%%%%%%%%%%%%%%%%%%%%%%%%
%
\section{Introduction}
A new era in particle physics has begun with the start-up of the CERN Large Hadron Collider (LHC).  With the unprecedented energies available, the observation of particles inaccessible to previous machines has become possible, as impressively proven by the discovery of a Higgs boson by the ATLAS~\cite{Aad:2012tfa} and CMS collaborations~\cite{Chatrchyan:2012xdj}. While this long-awaited observation advances our understanding of the mechanism responsible for electroweak symmetry breaking in the context of the Standard Model of elementary particles (SM) and many of its extensions, we are still left with a plethora of open questions that point towards the necessity of considering physics beyond the Standard Model (BSM).  Particularly strong indications for physics beyond the SM come from astrophysical observations that can only be accounted for by the existence of a large amount of Dark Matter (DM) in the universe. The SM, however, does not contain any particles that could serve as DM candidates with suitable properties. Thus, currently BSM scenarios featuring possible DM candidates are receiving increased attention (see, e.g., Ref.~\cite{Abercrombie:2015wmb} for a recent review). A particularly promising class of such models is comprised by supersymmetric theories postulating new particles that differ from their SM counterparts by their spin and acquire large masses by the mechanism of supersymmetry (SUSY) breaking. 
In the Minimal Supersymmetric Extension of the SM (MSSM) the conservation of R parity ensures that the lightest SUSY particle (LSP) is stable. For many parameter points of the MSSM the LSP is represented by a neutralino.
Being stable and electrically neutral, it provides an excellent candidate for fermionic DM. In the following, we will refer to neutralinos and charginos generically as electroweakinos or simply weakinos. 

In hadronic collisions, electroweakinos can be produced in pairs via electroweak (EW) interactions.  Because of the relatively small value of the EW coupling, the associated production cross sections are small. The ATLAS and CMS collaborations thus could only place relatively loose exclusion limits on the masses of these SUSY particles, in contrast to the much more severe limits available for the strongly interacting squarks and gluinos. These limits on the weakino masses are very model--dependent: For example, in a particular simplified scenario featuring weakino decays into sleptons, chargino and neutralino masses in the ranges $m_{\neu_1^\pm} \leq 700$~GeV and $m_{\neu_1^0} \leq 250$~GeV have been excluded, see Refs.~\cite{Khachatryan:2014qwa,Aad:2015jqa,Aad:2015eda,CMS:2016saj}.

The first calculation of the next-to-leading order (NLO) SUSY-QCD corrections to various electroweakino pair-production processes at hadron colliders was presented in Ref.~\cite{Beenakker:1999xh} and made available in the form of the public computer program \PROSPINO{}~\cite{Beenakker:1996ed}. Depending on the SUSY particle types and masses, NLO SUSY-QCD corrections of up to 45\% were reported for a collision energy of 14~TeV, resulting in a relative improvement of discovery limits for gauginos of about 10\% compared to the leading order (LO) estimate.
Complementary to the fixed-order calculation, transverse-momentum resummation effects to color-neutral gaugino pair-production processes were provided in Ref.~\cite{Debove:2009ia}, revealing the inadequacy of leading-order Monte Carlo simulations for a satisfactory description of transverse momentum spectra. Threshold resummation effects were addressed in Refs.~\cite{Li:2007ih,Debove:2010kf}, where a slight increase in invariant mass spectra and total cross sections, and a considerable stabilization with respect to the fixed-order predictions were found. A combination of transverse-momentum and threshold resummation effects was provided in Ref.~\cite{Debove:2011xj} and implemented in the \RESUMMINO{} code package~\cite{Fuks:2013vua}. 
One-loop EW corrections have been found to amount to about -6\% for representative parameter points for the associated production of a chargino and a neutralino at the LHC in Ref.~\cite{Hao:2006df}. 
SUSY-QCD corrections to neutralino-pair production in association with a jet were presented in Ref.~\cite{Cullen:2012eh}.

While total production cross sections for electroweakino pair-production processes can be obtained for a large variety of MSSM parameter points via the public computer program \PROSPINO{} or 
the  \RESUMMINO{} code package that additionally provides transverse-momentum and invariant-mass distributions of the gauginos, currently no dedicated Monte Carlo program exists for the calculation of differential distributions within arbitrary experimental selection cuts at NLO SUSY-QCD accuracy. Moreover, the afore-mentioned  public programs do not provide information on the gaugino decay chains, and cannot be interfaced easily with Monte-Carlo programs such as \PYTHIA{}~\cite{Sjostrand:2006za} for the simulation of parton-shower, underlying-event, and multi-parton interaction effects.
In principle, multi-purpose programs like \amc{}~\cite{Alwall:2014hca} do provide building blocks for the computation of arbitrary processes at NLO-QCD accuracy. In the context of SUSY processes with a complex resonance structure, though, human interaction is required for the subtraction of on-shell resonances that currently cannot be accounted for automatically in the \amc{} framework.

With the current work, we want to close existing gaps. We have developed a versatile code package for various electroweakino pair-production processes that provides NLO SUSY-QCD corrections to cross sections and differential distributions within arbitrary experimental selection cuts, and can be interfaced to parton shower programs via the \POWHEG{} matching procedure \cite{Nason:2004rx,Frixione:2007vw}. We are using the framework of the \POWHEGBOX{}~\cite{Alioli:2010xd}, a public repository for the simulation of scattering processes at hadron colliders at NLO-QCD accuracy matched with parton shower programs. For technical aspects related to features of the MSSM we build on experience gained in the implementation of slepton \cite{Jager:2012hd,Jager:2014aua} and squark pair-production processes \cite{Gavin:2013kga,Gavin:2014yga}  in the framework of the \POWHEGBOX{}.

In the following section, we will briefly describe technical aspects of our calculation that are specific to the implementation of electroweakino pair-production processes in the context of the \POWHEGBOX. In Sec.~\ref{sec:pheno} we will provide representative numerical results with a particular emphasis on the impact that NLO and parton-shower effects have on observables. Focusing on a specific SUSY benchmark point we will demonstrate how the code package we developed can be used for the calculation of experimentally accessible distributions including a simulation of supersymmetric decay chains. Our conclusions are given in Sec.~\ref{sec:conc}.  

%
%%%%%%%%%%%%%%%%%%%%%%%%%%%%%%%%%%%%%%%%%%%%%%%%%%%%%%%%%%%%%%%
%
\section{Framework of the calculation}
\label{sec:calc}
%
%%%%%%%%%%%%%%%%%%%%%%%%%%%%%%%%%%%%%%%%%%%%%%%%%%%%%%
Our implementation of weakino pair-production processes in the framework of the \POWHEGBOX{} builds on experience gained for related supersymmetric reactions, in particular slepton- and squark-pair production processes \cite{Jager:2012hd,Jager:2014aua,Gavin:2014yga,Gavin:2013kga}. Rather than going into general features required for the implementation of a new process in the  \POWHEGBOX{} repository, we here will focus on aspects that are specific to weakino pair production.

\begin{figure}[t]
\bec
\includegraphics[width=0.9\textwidth,clip]{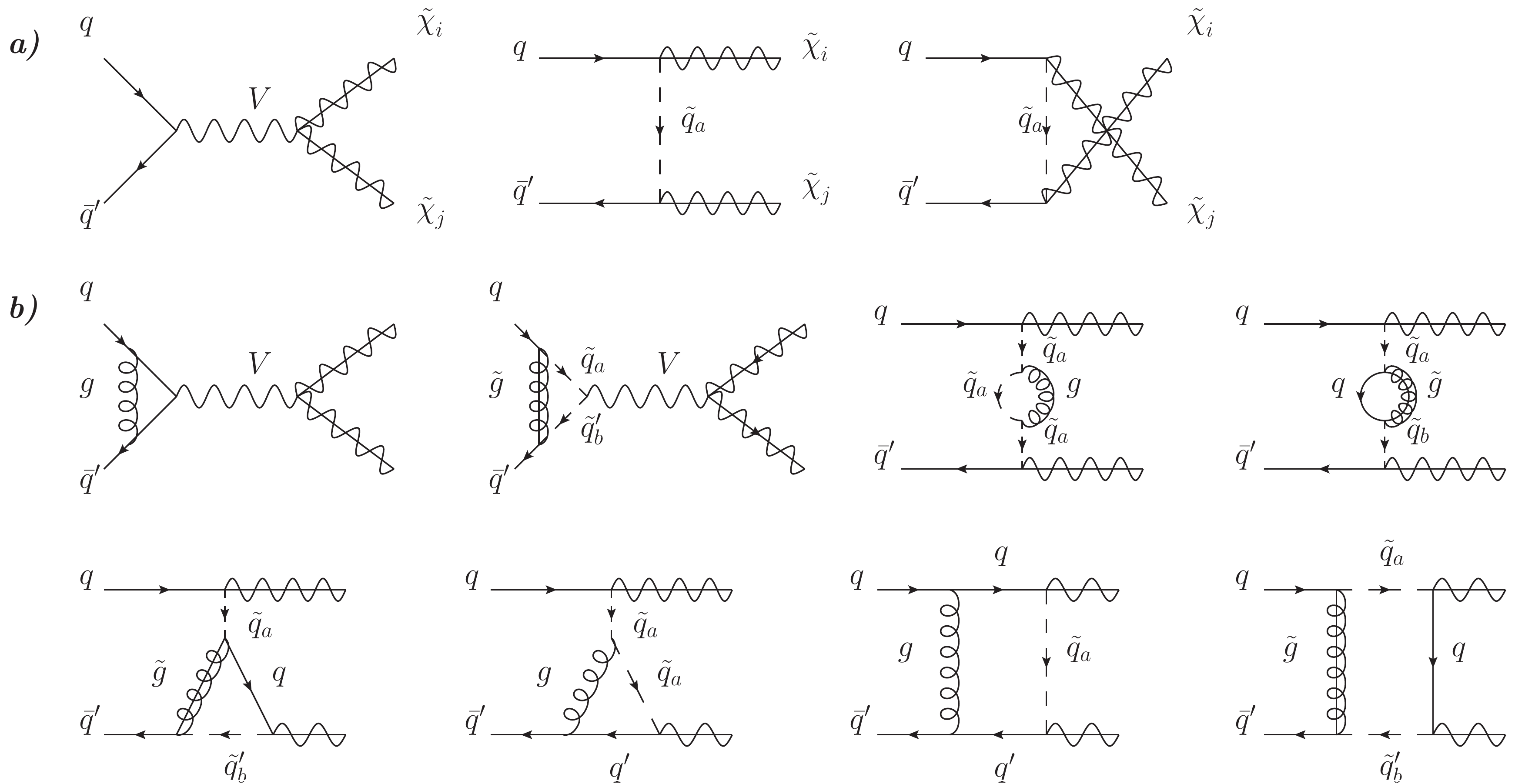}
\caption{Representative (a) tree-level and (b) one-loop diagrams for the production of a pair of weakinos at a hadron collider. Depending on the type of the produced weakinos, $V$ stands for $W_{}^\pm/Z/\gamma$,  and $a,b=1,2$.}
\label{fig:diagrams}
\eec
\end{figure}

At the leading order the production of a pair of weakinos proceeds via the tree-level diagrams presented in Fig.~\ref{fig:diagrams}~(a). In all channels the $s$--channel topology comprises Drell-Yan production, $q\bar{q}'\to V_{}^*$, followed by the splitting $V_{}^*\to \neu_i^{} \neu_j^{}$, where $\neu_i^{}$ stands for either a neutralino $\neu_i^0$ ($i=1\cdots 4$), or a chargino $\neu_i^\pm$ ($i=1,2$), depending on the process of interest. The vector boson $V$ denotes a $Z$~boson in the case of neutralino pair production, $V=W_{}^\pm$ for the production of a neutralino and a chargino, and $V=\gamma/Z$ for the production of a pair of charginos. In addition, diagrams with a squark being exchanged in the $t$-- or $u$--channel occur. In the case of the production of a chargino and a neutralino only either $t$-- or $u$--channel contributions arise, while for the other considered production processes both types of topologies contribute. We work in a scheme with five massless quark flavors in the initial state, i.e.\ $q/q' = u,d,s,c,b$. Numerically small bottom mass effects are disregarded throughout. This allows us to treat the scalar partners of these left-- and right--chiral fermions as mass eigenstates. The CKM matrix is taken to be diagonal. 

The NLO-QCD and SUSY-QCD corrections comprise virtual corrections to the $q\bar q'$-induced processes as well as real corrections with an extra parton in the final state. Only the sum of both corrections is infrared (IR) finite. Representative Feynman diagrams for the virtual corrections are shown in Fig.~\ref{fig:diagrams}~(b). They include vertex and box corrections with gluon, gluino, quark, or squark exchange, as well as self-energy corrections in the case of the $t$-- and $u$--channel diagrams with squark exchange. 
We use {\tt FeynArts 3.9}~\cite{Hahn:2000kx} to generate the virtual diagrams and {\tt FormCalc 8.4}~\cite{Hahn:1998yk} to calculate the amplitudes using the MSSM-CT model file of Ref.~\cite{Fritzsche:2013fta}. The scalar loop integrals~\cite{'tHooft:1978xw} are numerically calculated with {\tt LoopTools~2.12}~\cite{vanOldenborgh:1990yc,Hahn:1998yk}.
In order to cancel the ultraviolet (UV) divergences, a renormalization procedure has to be conducted. In practice, this requires the calculation of suitable counterterms.
We use the on-shell scheme for the renormalization of the wave functions of the massless incoming quarks as well as for the squark masses. Other fundamental parameters such as the electroweak coupling constant do not require renormalization at NLO in QCD. 

In order to regularize the UV divergent loop integrals, in principle there are two possibilities in a supersymmetric theory. The standard procedure of the \POWHEGBOX{} is the dimensional regularization scheme (DREG), where the entire calculation is done in $D=4-2\varepsilon$ dimensions. The same regularization procedure is used in most publicly available sets of parton distribution functions (PDFs) that are needed for the computation of cross sections at a hadron collider. However, this scheme breaks supersymmetry at the level of the gauge interactions by introducing a mismatch in the $(D-2)$ transverse degrees of freedom of the gauge bosons and the two degrees of freedom of the gauginos. 
Hence, while SUSY invariance requires that the quark-squark-weakino Yukawa coupling $\hat{g}$ and the associated $SU(2)$ gauge coupling $g$ be equal at all perturbative orders, this relation is violated in DREG. In order to remedy this deficiency, a finite SUSY restoring counterterm is added at next-to-leading order in the strong coupling $\alpha_s$~\cite{Martin:1993yx,Beenakker:1999xh,Hollik:2001cz}, 
\begin{align}
\hat{g} = g \left( 1 - \frac{\alpha_s^{}}{6\pi}\right).
\label{eq:susyrestore}
\end{align}
The expansion in $\alpha_s$ is done consistently to retain only the $\mathcal{O}(\alpha_s)$ term that is induced by this finite SUSY restoring counterterm in the amplitude squared.
%%%%
%
An alternative way of regularization is the dimensional reduction scheme (DRED) for the calculation of the finite part of the virtual corrections. In DRED fields remain defined in four dimensions, while loop momenta are defined in $D$ dimensions. Since this approach respects supersymmetry,  the SUSY-restoring counterterm of Eq.~(\ref{eq:susyrestore}) is not needed anymore. However, to comply with the intrinsic treatment of the IR~singularities in the \POWHEGBOX{}, a finite shift of the virtual amplitudes passed to the Monte-Carlo program is necessary~\cite{Alioli:2010xd}, 
\begin{align}
\mathcal{V} = \mathcal{V}_{}^{\rm DRED} - \frac43 \frac{\alpha_s^{}}{2\pi} \mathcal{B},
\end{align}
where $\mathcal{B}$ is the Born amplitude for a specific partonic subprocess calculated in four dimensions. As a cross-check, we have employed  both regularization methods in our calculation, and have in both schemes found the same value for the virtual amplitude $\mathcal{V}$ that enters the Monte-Carlo program.

\begin{figure}[t]
\bec
\includegraphics[scale=0.5,clip]{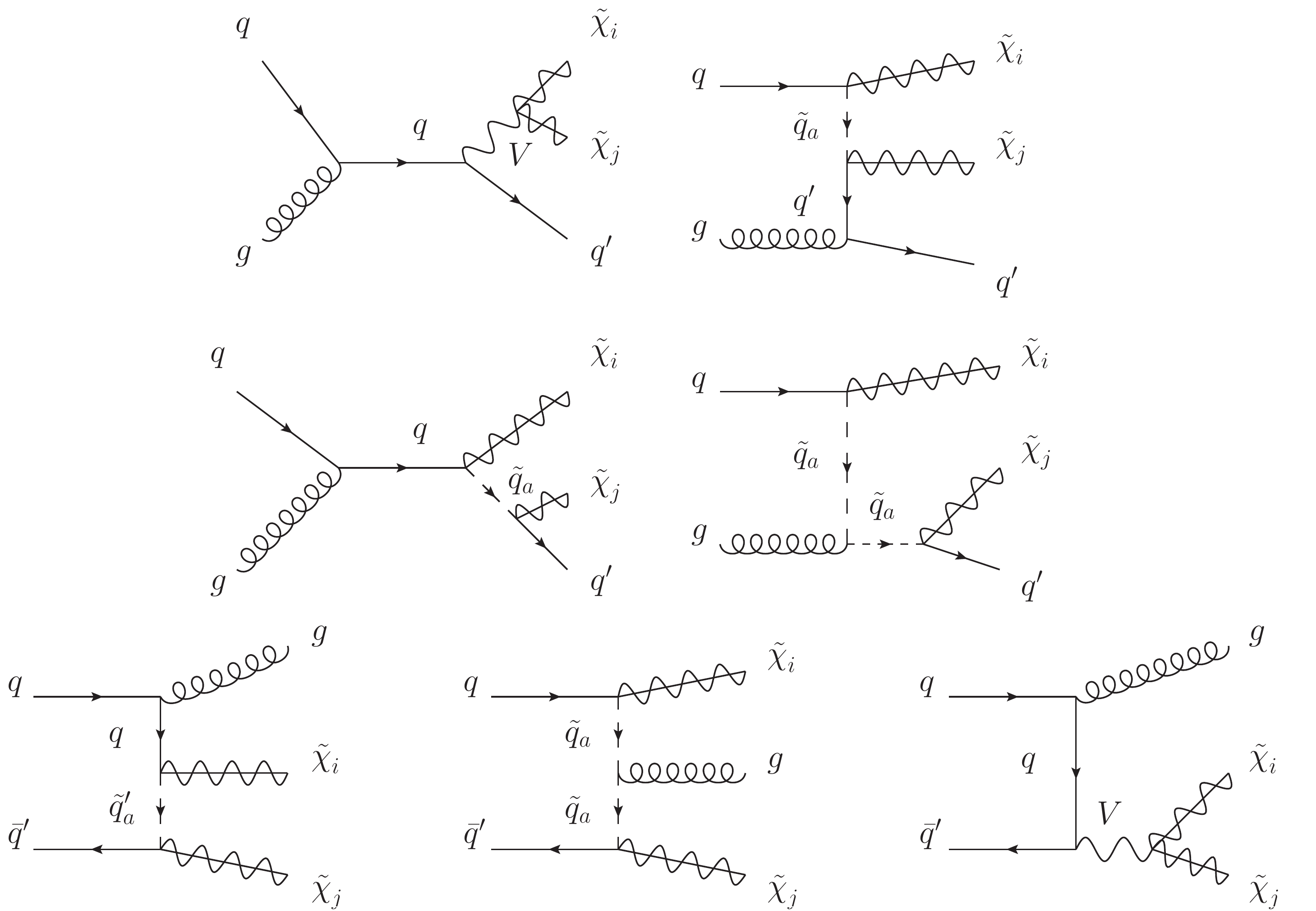}
\caption{Representative Feynman diagrams for the partonic subprocesses $q \bar{q}'\to \neu_i \neu_j g$ (lowest row), for diagrams including a squark resonance in the $q g\to \neu_i \neu_j q'$ channel (middle row), and for non-resonant diagrams in the $q g\to \neu_i \neu_j q'$ channel (upper row). In each case, $a=1,2$. }
\label{fig:diagrams_real}
\eec
\end{figure}

In order to calculate the real emission contributions and provide the ingredients necessary for the construction of IR subtraction terms by the \POWHEGBOX{},  we make use of a build tool based on {\tt MadGraph 4}~\cite{Murayama:1992gi,Stelzer:1994ta,Alwall:2007st}. It can be used to generate the Born, the color-- and spin--correlated Born and the real--emission amplitudes in a format that can be easily processed by the \POWHEGBOX{}. The IR divergences are canceled separately in the virtual and in the real parts by using the Frixione-Kunszt-Signer algorithm~\cite{Frixione:1995ms} that is implemented in the \POWHEGBOX{}. 
Representative real emission diagrams are displayed in Fig.~\ref{fig:diagrams_real}.

While the calculation of the real-emission contributions for the $q\bar q'$-induced subprocesses of type $q \bar q' \to \neu_i \neu_j g$ is straightforward, a subtlety arises in crossing-related partonic subprocesses with a quark in the final state. 
As noted in Ref.~\cite{Beenakker:1999xh},  subprocesses of the type $q g\to \neu_i \neu_j q'$ include two types of contributions: First, we encounter one-parton emission diagrams being part of the genuine NLO-QCD corrections to weakino pair production (representative diagrams are displayed in the upper row of Fig.~\ref{fig:diagrams_real}). Second, there occur  contributions that can be interpreted as tree-level diagrams for the on-shell production process $pp\to \tilde{q}_k \neu_i$, followed by the squark decay $\tilde{q}_k\to q' \neu_j$, if the squark is sufficiently heavy for a (quasi) on-shell decay into the respective weakino, i.e.\  $m_{\tilde{q}_k}>m_{\neu_j}$ (representative diagrams for this class of contributions are displayed in the middle row of Fig.~\ref{fig:diagrams_real}). 
This on-shell feature emerges not only in weakino pair production, but also in other supersymmetric production processes involving squarks or gluinos~\cite{Beenakker:1996ch,GoncalvesNetto:2012yt,Gavin:2013kga,Gavin:2014yga}, or in the case of $tW$ production in the framework of the SM~\cite{Re:2010bp}. While the genuine real-emission contributions of the first type clearly have to be taken into account in our NLO~calculation, the resonant contributions that are to be considered part of the different production process $pp\to \tilde{q}_k \neu_i$ need to be removed consistently in order to avoid double-counting. 
To this end, we make use of a scheme that in a slightly different variant has first been applied in \PROSPINO{}~\cite{Beenakker:1996ch}, and more recently has been adapted for the code structure of the \POWHEGBOX{} in Refs.~\cite{Re:2010bp,Gavin:2013kga}.  More specifically, we extend the procedure developed for the related case of squark pair production in the \POWHEGBOX{}~\cite{Gavin:2013kga,Gavin:2014yga} to the richer resonance structure of weakino pair-production processes, as more diagrams are involved in that case.
%
%%%%%%
%
%%%%%

Each resonant diagram occurring in a subprocess of type $q g\to \neu_i \neu_j q'$ exhibits a propagator that diverges when the intermediate squark goes on shell. For instance, in the case of a $\tilde{q} \to \neu_j q'$ decay,  this implies $(p_{\neu_j} + p_{q'})^2\to m_{\tilde{q}}^2$ in terms of the momenta of the external particles. Such would-be divergencies can be tamed by assigning a finite width $\Gamma_{\tilde{q}}$ to the respective propagator, 
\beq
\frac{1}{(p_{\neu_j} + p_{q'})^2 - m_{\tilde{q}}^2}
\to
\frac{1}{(p_{\neu_j} + p_{q'})^2 - m_{\tilde{q}}^2 + i m_{\tilde{q}} \Gamma_{\tilde{q}}},
\eeq
as is done by default in our  {\tt MadGraph}-based real-emission amplitudes.  

Having determined the resonance structure of the resonant diagrams that are considered as part of a squark-weakino rather than a weakino pair-production process, we are now in a position to devise an on-shell (OS) subtraction procedure for isolating the genuine weakino pair-production process of interest.   
To this end, we split the full real-emission amplitude into a purely resonant contribution, $\mc{M}_\mr{res}$, and a regular remainder, $\mc{M}_\mr{reg}$. Specifically, in the case of neutralino pair production, eight resonant diagrams $\mc{M}_\mr{res}^k$ with $k=1,\ldots, 8$ contribute to $\mc{M}_\mr{res}$, while for production processes involving charginos, only four resonant diagrams occur. 
In order to remove the resonant squark contributions, for each resonant diagram squared $\left|\mc{M}_\mr{res}^k\right|_{}^2$ we introduce a counterterm of the form 
\bea
\left| \mc{M}_\mr{res}^{k, \rm CT}(\Gamma_{\tilde{q}})\right|^2 
&=&
\Theta\left(\hat{s}-(m_{\tilde{q}}+m_{\chi_i})^2\right) \Theta\left(m_{\tilde{q}}-m_{q'}-m_{\chi_j}\right)
\non\\
&&\times
\frac{m_{\tilde{q}}^2 \Gamma_{\tilde{q}}^2
}{
(p_\mr{res}^2-m_{\tilde{q}}^2)^2 + m_{\tilde{q}}^2  \Gamma_{\tilde{q}}^2
}
\left| \mc{M}_\mr{res}^k (\Gamma_{\tilde{q}})\right|^2_\mr{remapped} \,,
\label{eq:onshell}
\eea
where $p_\mr{res}=(p_{\neu_j} + p_{q'})$, and the momenta entering  $\mc{M}_\mr{res}^k$ are to be remapped to the on-shell kinematics, c.f.~Ref.\cite{GoncalvesNetto:2012yt}. The first step function in this equation  ensures that the partonic center-of-mass energy $\hat{s}$  is sufficient to generate both an on-shell intermediate squark and an on-shell spectator~$\neu_i$. The second theta-function guarantees that the squark has a mass larger than the sum of the masses of the two particles into which it decays, so that it can become on-shell. Since we only consider massless quarks, $m_{q'} = 0$~GeV in our calculation. We also stress that the decay width $\Gamma_{\tilde{q}}$ introduced in Eq.~(\ref{eq:onshell}) is to be viewed as a technical regulator in the on-shell subtraction procedure.  It may, but not necessarily has to, be identified with the actual physical  
decay width for the resonant squark. Since after the on-shell subtraction results should not depend on the resonant contributions, final results must be independent of $\Gamma_{\tilde{q}}$.

After having identified the counterterms needed for the on-shell subtraction procedure, 
we are in a position to perform the phase-space integration separately for the regular and the on-shell subtracted resonant contributions, 
\begin{align}
\sigma_{\rm real} &= \sigma_{\rm real}^{\rm reg} + \sigma_{\rm real}^{\rm OS},
\end{align}
with
\begin{align}
        \sigma_{\rm real}^{\rm reg} &= \int d\Phi_{3} \left|\mc{M}_{\rm reg}\right|^2, \\
        \sigma_{\rm real}^{\rm OS} &= \sum\limits_{k}\int d\Phi_{3}^{\rm OS}
        \left[
        \left| \mc{M}_\mr{res}^k(\Gamma_{\tilde{q}})\right|^2 
        - \mc{J}_k^{}\left| \mc{M}_\mr{res}^{k,\rm CT}(\Gamma_{\tilde{q}})\right|^2
        \right]\,. 
\end{align}
While the regular contributions are to be evaluated for the standard real-emission kinematics, 
the counter-term contributions have to be integrated over a remapped phase-space that is obtained  from the original three-body phase space $d\Phi_{3}$ via a Jacobian factor $\mc{J}_k^{}$,
\begin{align}
	  \mc{J}_k^{} = \frac{s_{\tilde{q}}}{m_{\tilde{q}}^2} \frac{\lambda^{1/2}(m_{\tilde{q}}^2,\hat{s},m_{\neu_i}^2) \,\lambda^{1/2}(m_{\tilde{q}}^2,m_{\neu_j}^2,m_{q'}^2)}{\lambda^{1/2}(s_{\tilde{q}},\hat{s},m_{\neu_i}^2)\, \lambda^{1/2}(s_{\tilde{q}},m_{\neu_j}^2,m_{q'}^2)}\,,
\end{align} 
where $s_{\tilde{q}} = p_\mr{res}^2$, and $\lambda$ denotes the Kaellen-function, 
\begin{align}
        \lambda(x,y,z) =& \,x^2 + y^2 + z^2 - 2(x\,y + y\,z + z\,x)\,.
\end{align}
If the limits of the phase-space integration would not be adapted appropriately, an integration over the entire three-body phase-space would combine on-shell and off-shell contributions inconsistently. For further details on the rescaling of phase-space, see Refs.~\cite{Gavin:2013kga,Gavin:2014yga}.
For the actual evaluation of  $\sigma_{\rm real}^{\rm OS}$ in the \POWHEGBOX{}, we have devised a routine allowing for a mapping of phase space according to a specific resonance structure $k$.  Following this procedure, we can in principle handle an arbitrary number of resonance structures. The routine we developed could thus be used for future \POWHEGBOX{} implementations of other processes requiring an on-shell resonance subtraction. We note that the on-shell subtraction procedure we are using has the advantage of numerical stability, but violates gauge invariance as $\Gamma_{\tilde{q}}\neq 0$. In order to overcome this drawback alternative methods have been explored in the literature~\cite{Gavin:2013kga}, but were found to exhibit other disadvantages such as the requirement of artificial cuts, and also being quite  involved in a {\tt MadGraph}-based implementation of the real corrections. However, gauge invariance violating contributions in the approach we are using are numerically negligible, as demonstrated by the independence of our results on the technical parameter $\Gamma_{\tilde{q}}$ discussed below. 

\begin{figure}[t]
\bec
\includegraphics[width=0.70\textwidth]{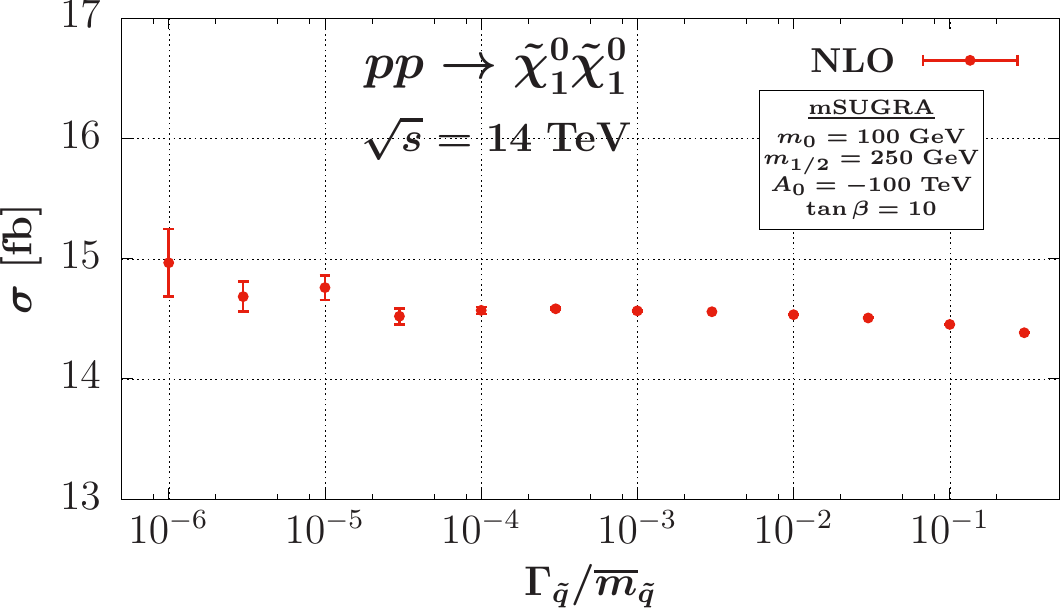}
\caption{Dependence of the total cross section for the process $pp\to \neu_1^0\neu_1^0$ with $\sqrt{s}=14$~TeV on the regulator $\Gamma_{\tilde{q}}$ normalized to the average squark mass of $\overline{m}_{\tilde{q}} = 556$~GeV.}
\label{fig:reg-dep}
\eec
\end{figure}

In order to verify the validity of our implementation, we have performed a number of checks. 
First, we have tested that, after the subtraction of on-shell resonances, for collinear momentum configurations real-emission and IR subtraction terms approach each other.  
Second, we have found that the dependence of our predictions for weakino pair-production cross sections on the technical regulator $\Gamma_{\tilde{q}}$ is negligible. Figure~\ref{fig:reg-dep}  illustrates the regulator dependence of the neutralino pair-production cross section for a SUSY benchmark point that features squarks heavy enough to on-shell decay into a neutralino and a quark. We consider the mSUGRA spectrum SPS~1a~\cite{Allanach:2002nj}  with $m_0^{} = 100$ GeV, $m_{1/2}^{}  = 250$ GeV, $A_0 = -100$ GeV, $\mr{sgn}(\mu) =  +1$, and $\tan\beta=10$ at the GUT scale, resulting in the lightest neutralino mass $m_{\neu_1^0}^{} = 96.69$ GeV and the first--generation squark masses $m_{\tilde{u}_L/\tilde{u}_R/\tilde{d}_L/\tilde{d}_R}^{} = 561.1/549.3/568.4/545.2$ GeV. 
Although this benchmark point is already excluded by experiment, see, for example, Ref.~\cite{AbdusSalam:2011fc}, we use it in order to illustrate the technical details of the regulator dependence as it easily provides a spectrum for which the squark masses induce resonances to be regulated. We do not use this benchmark point for phenomenological studies.
In the range  $\Gamma_{\tilde{q}}/\overline{m}_{\tilde{q}} = 10^{-5}\;\mr{to}\;10^{-1}$, where $\overline{m}_{\tilde{q}} = 556$ GeV is the average of the four squark masses of the first generation, the dependence of the cross section on the regulator is entirely negligible, thus confirming the stability of the applied on-shell subtraction procedure. 
Finally, we have computed total cross sections at LO and NLO accuracy in the setup of Ref.~\cite{Beenakker:1999xh} and found agreement with the published results. 

%
%%%%%%%%%%%%%%%%%%%%%%%%%%%%%%%%%%%%%%%%%%%%%%%%%%%%%%%%%
%
%
%%%%%%%%%%%%%%%%%%%%%%%%%%%%%%%%%%%%%%%%%%%%%%%%%%%%%%%%%%
%

%
%%%%%%%%%%%%%%%%%%%%%%%%%%%%%%%%%%%%%%%%%%%%%%%%%%%%%%%%%%
%
%%%%%%%%%%%%%%%%%%%%%%%%%%%%%%%%%%%%%%%%%%%%%%%%%%%%%%%%%%%%%%%
%
\section{Phenomenological results}
\label{sec:pheno}
A collection of electroweakino pair-production processes  
will be made publicly available in the framework of the \POWHEGBOX{} via the project website {\tt \url{http://powhegbox.mib.infn.it/}}. Since the public code can be used for specific user applications, we refrain from presenting an extensive numerical analysis here, but only intend to highlight some representative phenomenological results. 

For our numerical studies, we consider proton-proton collisions at the LHC with a center-of-mass energy of $\sqrt{s}=14$~TeV. For the parton distribution functions (PDFs) of the proton we use the PDF4LHC15 NLO set~\cite{Butterworth:2015oua} as implemented in the LHAPDF library~\cite{Buckley:2014ana}. Since no LO set is provided by the PDF4LHC15 working group, we use the NLO set also for the computation of LO results.
Unless explicitly specified otherwise, we choose fixed values for the renormalization and factorization scales, $\mur$ and $\muf$, proportional to the sum of the masses of the weakinos $\neu_A$ and $\neu_B$  produced in the specific process under consideration, $\mur=\muf=\xi\mu_0 $ with $\mu_0 = m_{\neu_A}+m_{\neu_B}$. The scale parameter $\xi$ is set to one by default. When combining fixed-order results with a parton-shower program, we use \PYTHIA{}~6.4.25~\cite{Sjostrand:2006za}. QED radiation, underlying event, and hadronization effects are switched off throughout. Partons arising from the real-emission contributions of the NLO-QCD calculation or from the parton shower are recombined into jets according to the anti-$k_T$ algorithm \cite{Cacciari:2008gp} as implemented in the {\tt FASTJET} package \cite{Cacciari:2011ma} with $R=0.4$ and $\left|\eta^\mr{jet}\right|<4.5$. 

As electroweak input parameters we choose the $Z$ boson mass, $m_Z^{}=91.1876$~GeV, the electromagnetic coupling, $\alpha^{-1}_{}(m_Z)=127.934$, and the Fermi constant, $G_F^{}=1.16638\times 10^{-5}_{}$~GeV$_{}^{-2}$. The other SM and MSSM parameters required for our calculations are provided in the form of a file complying with the SUSY Les Houches Accord (SLHA)~\cite{Skands:2003cj,Allanach:2008qq} that can be computed with an independent external spectrum calculator. We have used the {\tt SuSpect 2.43}  program~\cite{Djouadi:2002ze} for the calculation of the spectrum and the {\tt SDECAY} program~\cite{Muhlleitner:2003vg} for the decay widths and branching fractions to obtain such an SLHA file.  Specifically, we consider a minimal supergravity (mSUGRA) benchmark point suggested in Ref.~\cite{Francescone:2014pza} that is consistent with a Higgs mass of about 126~GeV as well as further collider and dark matter constraints. This benchmark point is characterized by the following SUSY input parameters: $m_{1/2}^{} = 470$~GeV, $m_0^{} = 6183$~GeV, $A_0^{} = -4469$~GeV, $\tan\beta = 52.1$, $\mr{sgn}(\mu) = +1$.
These are resulting in neutralino masses of
\bea
m_{\neu_1^0} = 207.0~\GeV, \ \ \ \
m_{\neu_2^0} = 405.9~\GeV, \ \ \ \
m_{\neu_3^0} = 598.1~\GeV, \ \ \ \
m_{\neu_4^0} = 612.9~\GeV,\eea
and chargino masses of 
\bea
m_{\neu_1^\pm} = 405.8~\GeV, \ \ \ \
m_{\neu_2^\pm} = 613.2~\GeV\,.
\eea
The squark masses are equal for the first and second generation, but different for the third generation. The numerical values are as follows:
\bea
m_{\dsql/\ssql} = 6.172~\TeV, \ \ 
m_{\dsqr/\ssqr} = 6.193~\TeV, \ \ 
m_{\usql/\csql} = 6.172~\TeV, \ \ 
m_{\usqr/\csqr} = 6.190~\TeV\,,\nonumber
\eea
\vspace{-10mm}
\bea
m_{\bsql} = 4.132~\TeV, \ \ 
m_{\bsqr} = 4.591~\TeV, \ \ 
m_{\tsql} = 3.577~\TeV, \ \ 
m_{\tsqr} = 4.112~\TeV\,.
\eea

\begin{figure}[t]
\bec
\includegraphics[width=0.70\textwidth]{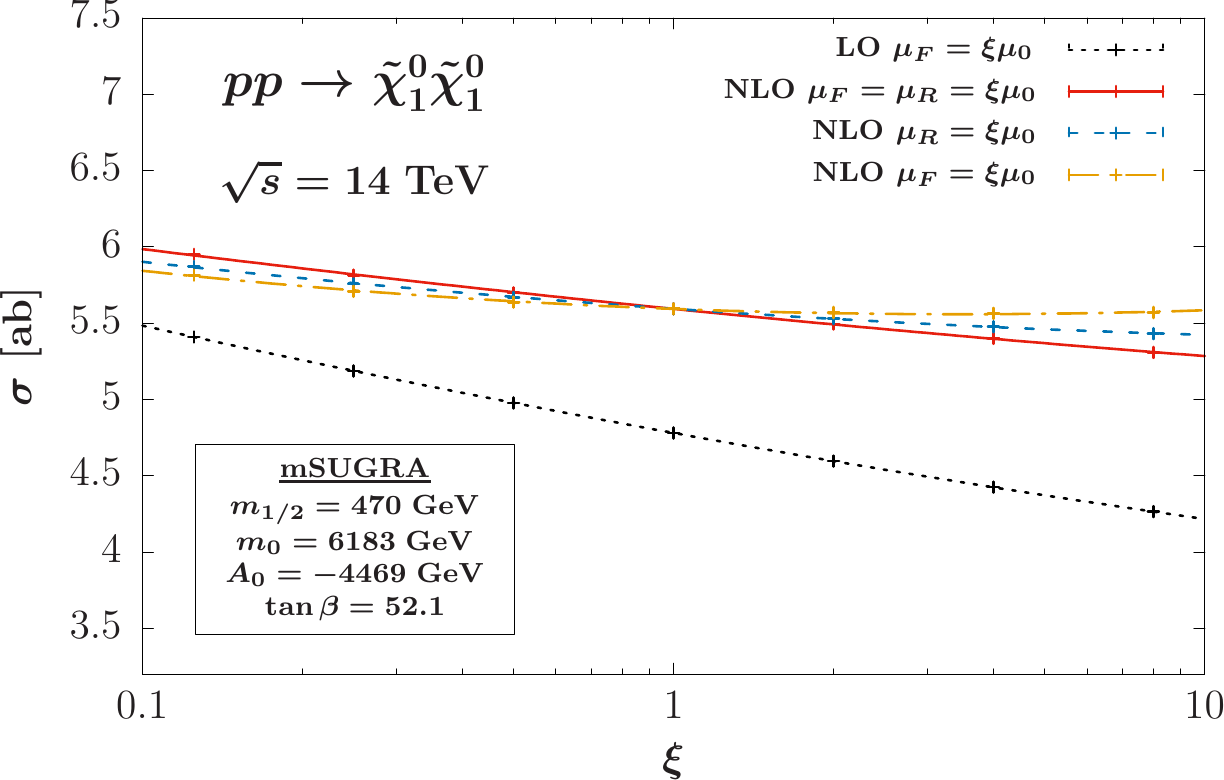}
\caption{
Dependence of the total cross section for the process $pp\to \neu_1^0\neu_1^0$ with $\sqrt{s}=14$~TeV on the factorization and renormalization scales. The NLO curves show the cross section as a function of the scale parameter $\xi$ for three different cases: $\mur=\muf=\xi \mu_0$ (solid red line), $\mur=\xi \mu_0, \muf=\mu_0$ (dashed blue line), and $\mur=\mu_0, \muf=\xi \mu_0$ (dot-dashed yellow line). The LO cross section only depends on $\muf=\xi \mu_0$ (dotted black line). In each case, $\mu_0=2\, m_{\neu_1^0}$.
}
\label{fig:scale-dep}
\eec
\end{figure}

For this benchmark point, we first consider the neutralino pair-production process $pp\to \neu_1^0 \neu_1^0$. We find a total cross section of $\sigma^\mr{LO} = 4.780$~ab at LO and of  $\sigma^\mr{NLO} =  5.595$~ab at NLO. The NLO SUSY-QCD corrections thus enhance the production rate by more than 15\%.
In order to quantify the dependence of these results on the unphysical renormalization and factorization scales, we have varied $\mur$ and $\muf$ in the range $0.1\mu_0$ to $10\mu_0$ around our default choice $\mur=\muf=\mu_0=2\, m_{\neu_1^0}$, c.f.\ Fig.~\ref{fig:scale-dep}. At LO, neutralino-pair production is a purely electroweak process and thus only depends on $\muf$ via the parton distribution functions of the scattering protons. The scale behavior of the LO results directly reflects the $\muf$ dependence of the (anti-)quark distribution functions in the probed kinematic regime. At NLO, additionally $\mur$ enters and, in contrast to $\muf$ being effectively accounted for only at lowest order, dominates the scale uncertainty of $\sigma^\mr{NLO} $. However, in the typically considered range $\mu_0/2$ to $2\mu_0$ the NLO cross section changes by only about 3\%, indicating that the perturbative expansion is rather stable, and the scale uncertainty is reduced compared to the LO predictions.
%
%%%%%%%%%%%%%%%%%%%%%%%%%%%%%%%%%%%%%%%%%%%%%%%%%%%%%%%%%%
%
%
%%%%%%%%%%%%%%%%%%%%%%%%%%%%%%%%%%%%%%%%%%%%%%%%%%%%%%%%%%
%

\begin{figure}[t]
\bec
\includegraphics[width=0.49\textwidth]{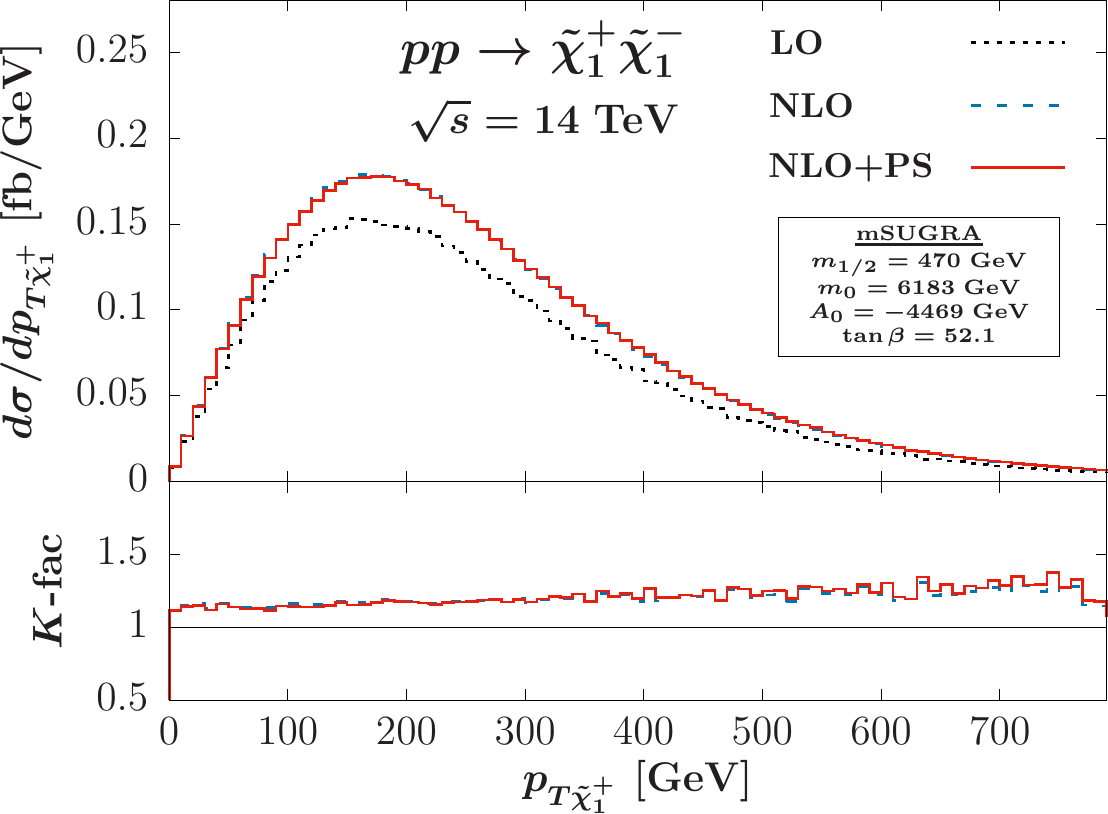}\hspace{2mm}
\includegraphics[width=0.48\textwidth]{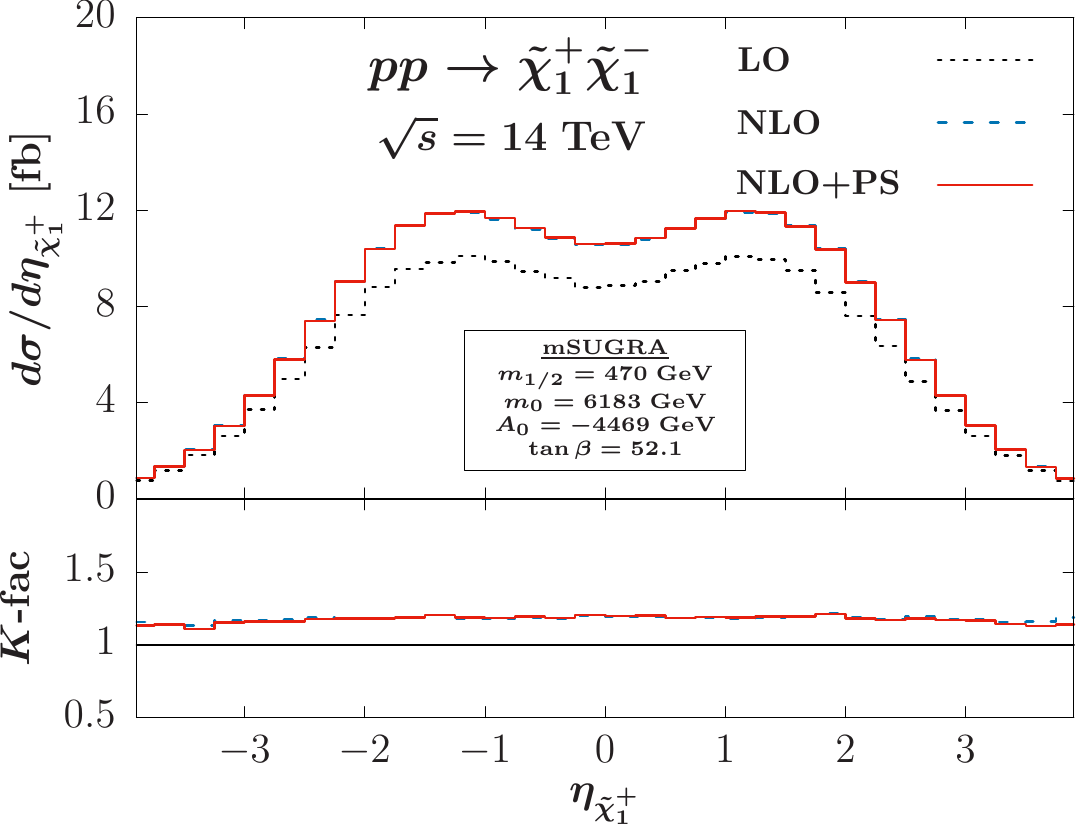}
\caption{
Transverse-momentum (left) and pseudorapidity distribution (right) of the $\neu_1^+$ in the process $pp\to \neu_1^+\neu_1^-$ at LO (dotted black lines), NLO (dashed blue lines), and \NLOPS{} (solid red lines) for our default setup. 
}
\label{fig:pt_eta_A_x1x1}
\eec
\end{figure}

\begin{figure}[t]
\bec
\includegraphics[width=0.49\textwidth]{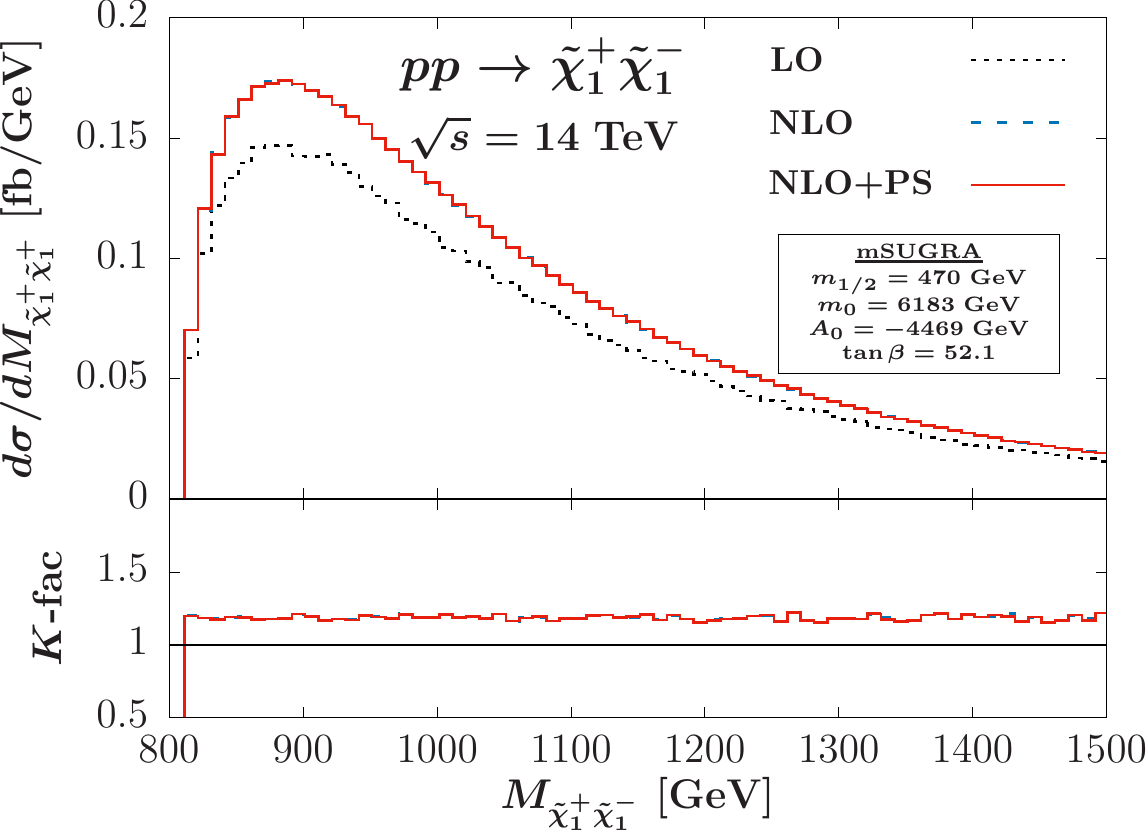}
\includegraphics[width=0.485\textwidth]{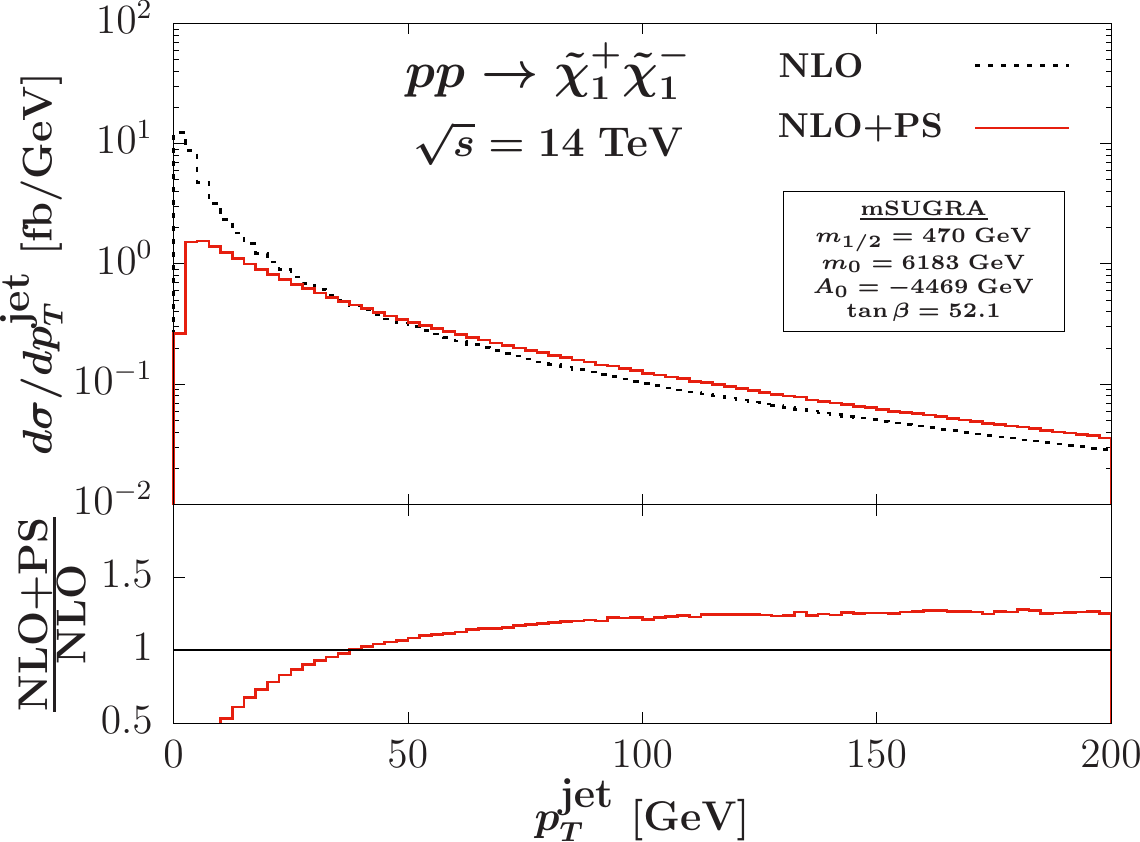}
\caption{
Invariant-mass distribution of the chargino-pair (left) and transverse-momentum distribution of the hardest jet (right) for the process $pp\to \neu_1^+\neu_1^-$ at LO (dotted black lines), NLO (dashed blue lines), and \NLOPS{} (solid red lines) for our default setup. 
}
\label{fig:m_jet_x1x1}
\eec
\end{figure}

In order to assess the impact of the higher-order corrections and parton shower effects on kinematic features of weakino pair production, we consider the representative chargino pair-production process $pp\to\neu_1^+ \neu_1^-$. Numerical uncertainties are at the permille level and not shown in the plots that follow.
Figure~\ref{fig:pt_eta_A_x1x1} illustrates the transverse-momentum and pseudo-rapidity  distributions of the $\neu_1^+$  at fixed order, and after the matching of the NLO result to the parton shower (\NLOPS). Analogous results are obtained for the other chargino, $\neu_1^-$. 
In Fig.~\ref{fig:m_jet_x1x1}~(left) we depict the invariant-mass distribution of the chargino pair. As expected from the above discussion of total cross sections for the related case of neutralino-pair production, we notice that the normalization of these distributions changes significantly when going from LO to NLO. On the other hand, their shapes are only slightly affected by the NLO corrections, as illustrated by the dynamical $K$-factors, 
\beq
K=\frac{d\sigma^\mr{NLO}}{d\sigma^\mr{LO}}\,, 
\eeq
which turn out to be mostly flat.
Obviously, parton-shower effects on the massive final state are very small for all considered distributions, which is largely due to the large mass and color-neutral nature of the supersymmetric final state. 
Details of the parton-shower settings will thus barely affect predictions for observables related to the charginos at \NLOPS{} level. Because of the small impact parton shower effects have on NLO results, in the figures the NLO and \NLOPS{} curves are almost indistinguishable.
%
%%%%%%%%%%%%%%%%%%%%%%%%%%%%%%%%%%%%%%%%%%%%%%%%%%%%%%%%%%
%
%
%%%%%%%%%%%%%%%%%%%%%%%%%%%%%%%%%%%%%%%%%%%%%%%%%%%%%%%%%%

More pronounced effects of the parton shower emerge in jet observables, such as the transverse-momentum distribution of the hardest jet shown in Fig.~\ref{fig:m_jet_x1x1}~(right). For the reaction $pp\to\neu_1^+ \neu_1^-$, at NLO, jets can only result from a hard parton of the real-emission contributions. After matching with a parton shower, additional jets can occur that will, however, be mostly soft or collinear. From the displayed figure it is apparent that while in the fixed-order calculation the transverse-momentum distribution of the jet diverges towards small values of $p_T^\mr{jet}$, the Sudakov form factor of the \NLOPS{} calculation tames this would-be divergence. We note, however, that a precise description of jet observables in weakino pair-production processes would require considering the related reactions with an associated jet being present at LO already. Only a full NLO calculation for the  $\neu_1^+ \neu_1^-+\mr{jet}$ production process would yield NLO-accurate predictions for jet distributions. In our calculation of  $pp\to\neu_1^+ \neu_1^-$, jet observables are described effectively only at LO accuracy and thus associated with significant theoretical uncertainties.
Our results confirm the findings obtained in the context of a jet veto resummation formalism \cite{Tackmann:2016jyb} for the related case of slepton pair production,  which revealed that theoretical uncertainties at the lowest resummation order are large enough to weaken current exclusion limits relying on searches making use of jet vetoes.
%
%%%%%%%%%%%%%%%%%%%%%%%%%%%%%%%%%%%%%%%%%%%%%%%%%%%%%%%%%%
%
%
%%%%%%%%%%%%%%%%%%%%%%%%%%%%%%%%%%%%%%%%%%%%%%%%%%%%%%%%%%
%

%
%%%%%%%%%%%%%%%%%%%%%%%%%%%%%%%%%%%%%%%%%%%%%%%%%%%%%%%%%%
%
\begin{figure}[t]
\bec
\includegraphics[width=0.35\textwidth]{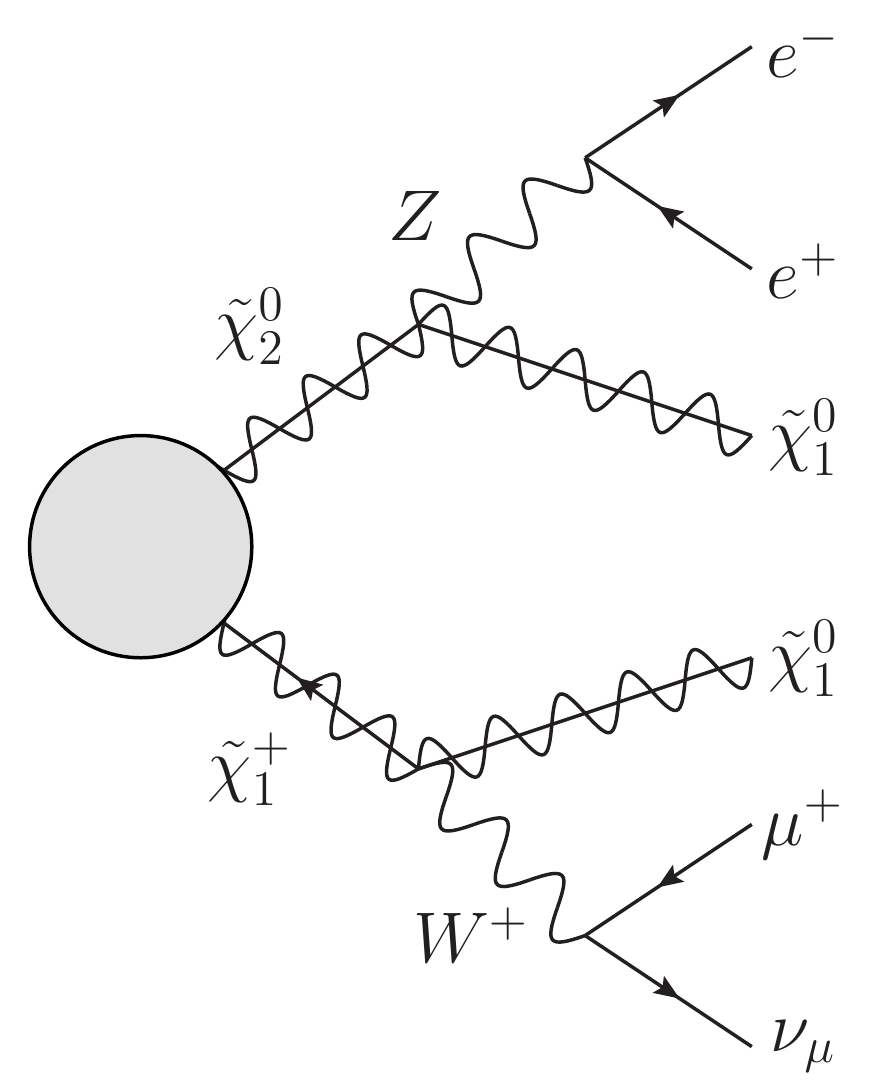}
\caption{Leptonic decay chain for the process $pp\to\neu_2^0 \neu_1^+$, giving rise to an $e^-e^+ \mu^++E_T^\mr{miss}$ final state.}
\label{fig:decay-chain}
\eec
\end{figure}
%
%%%%%%%%%%%%%%%%%%%%%%%%%%%%%%%%%%%%%%%%%%%%%%%%%%%%%%%%%%
%

In many SUSY scenarios, the $\neu_1^0$ represents the LSP that, due to the requirement of R-parity conservation, does not decay. Being electrically neutral, such an LSP cannot be observed directly in the detector, but only via its imprint on the missing transverse-energy spectrum. Depending on the mass hierarchy of a SUSY parameter point, heavier neutralinos and charginos decay via chains into a combination of stable particles, such as partons, leptons, neutrinos, and the LSP. Particularly clean experimental signatures emerge from final states with charged leptons that are rare in the context of the Standard Model. A prime example is provided by the leptonic decay chain of the process $pp\to\neu_2^0 \neu_1^+$ that gives rise to a three-lepton final state as depicted in Fig.~\ref{fig:decay-chain}. Having full access to supersymmetric decay chains in a Monte-Carlo simulation is thus of great phenomenological relevance. 
The codes we developed for weakino pair-production processes offer such an option by an interface to  the SUSY decay feature of \PYTHIA{}. We can thus provide predictions that are at the same time NLO accurate for the hard weakino pair-production process, include parton-shower emission effects, and give full access to the kinematic properties of the stable particles in specific decay chains using the narrow-width approximation.

\begin{figure}[t]
\bec
\includegraphics[width=0.49\textwidth]{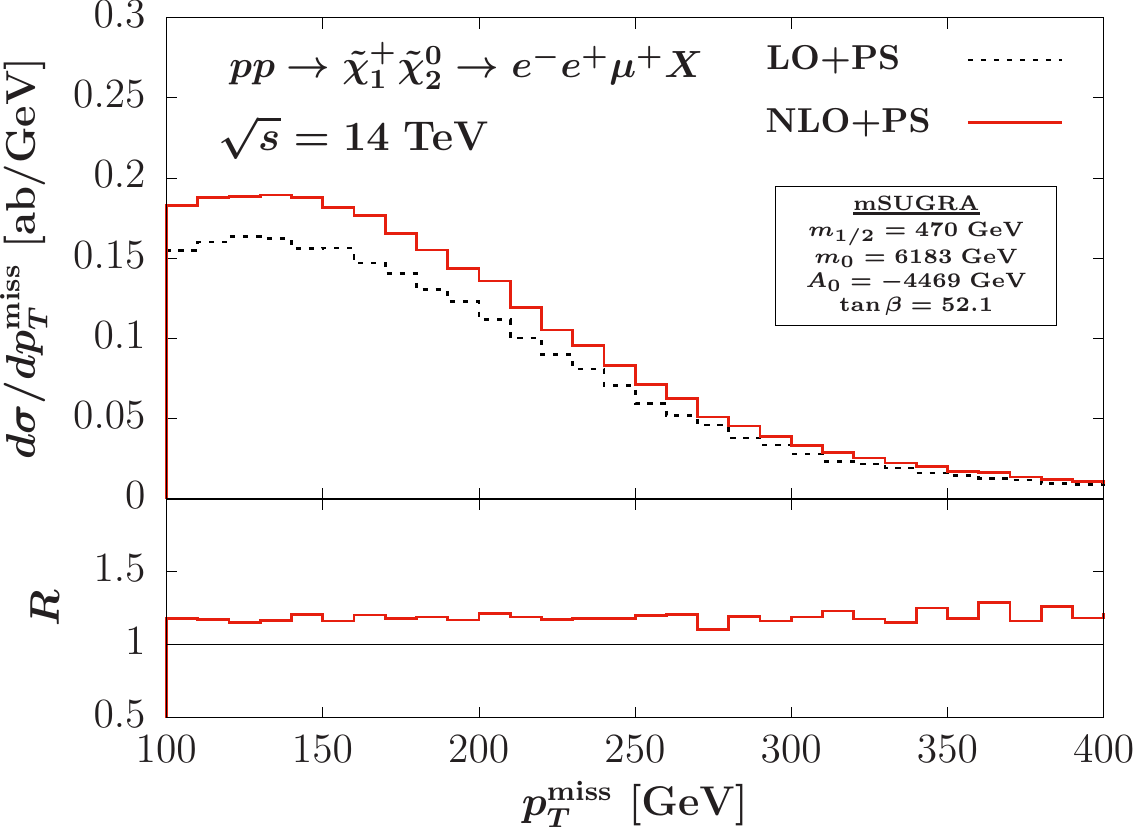}\hspace{2mm}
\includegraphics[width=0.48\textwidth]{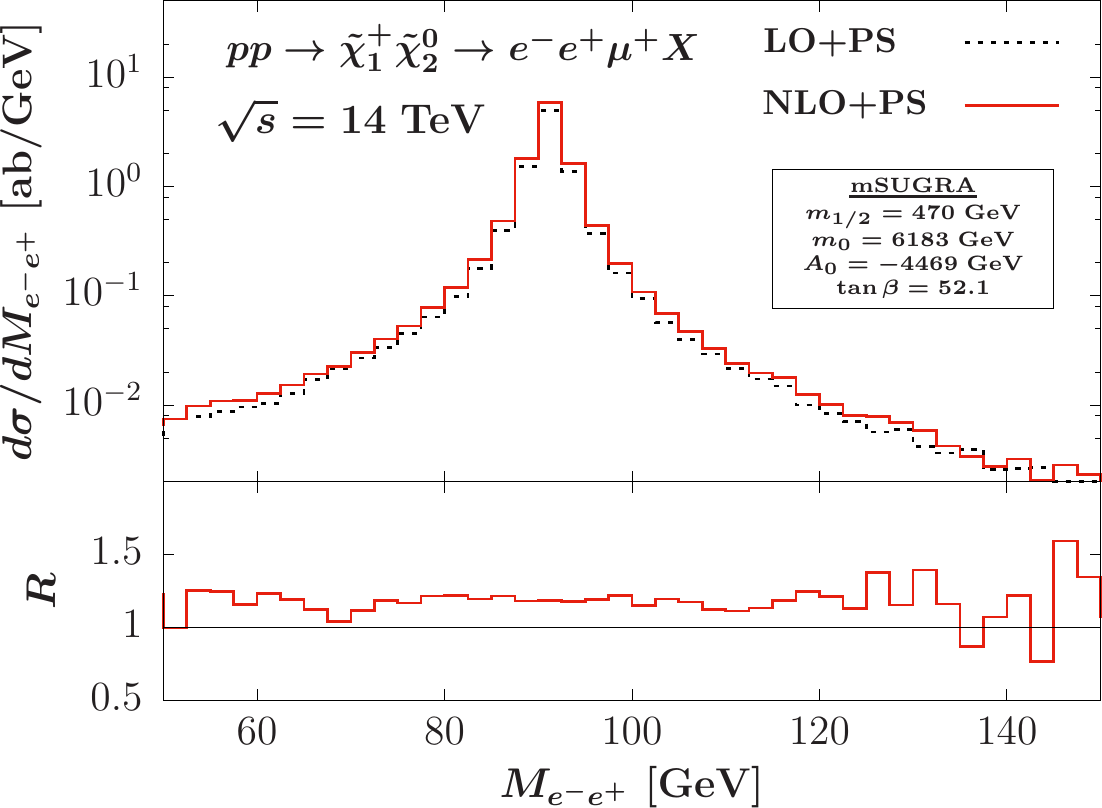}
\caption{
Missing transverse momentum (left) and invariant mass distribution of the $e^+e^-$ system (right) emerging in the $e^+e^-\mu^+ + E_T^\mr{miss}$  
decay mode of the process $pp\to \neu_2^0\neu_1^-$ at LO (dotted black lines) and NLO (solid red lines) matched with \PYTHIA{}, after the leptonic cuts of Eqs.~(\ref{eq:lcuts}) and (\ref{eq:ptmiss-cut}) are applied.
}
\label{fig:ptmiss_n2x1}
\eec
\end{figure}

To illustrate this feature, we focus on final states with three charged leptons plus missing transverse energy arising from the $\neu_2^0 \neu_1^+$ production process. For the setup of this simulation, we follow closely the strategy of the ATLAS analysis reported in Ref.~\cite{Aad:2015jqa}. We only consider events with an electron, a positron, a muon, and a large amount of missing transverse energy in the final state. Each charged lepton is required to exhibit non-vanishing transverse momentum, be located in the central-rapidity region and sufficiently well separated from each other in the rapidity-azimuthal angle plane,
\beq
\label{eq:lcuts}
p_T^\ell > 10~\GeV\,,\quad 
|\eta^\ell | < 2.5\,,
\quad \Delta R(\ell, \ell') > 0.05\,.
\eeq
In addition, the missing transverse momentum is required to be large,
\beq
\label{eq:ptmiss-cut}
p_T^\mr{miss} > 100~\GeV\,.
\eeq
This latter observable is computed from the negative sum of the final-state particles that are detected, i.e. the electron, positron, muon, and jets with a transverse momentum $p_T^{j} \geq 20$~GeV, similar to what is done in the experimental analyses.
As the sum of the transverse momenta of the final-state particles should add to zero, this is effectively similar to the sum over the non-detected particles, i.e. the LSP,  the neutrinos emerging in the decay chain, and the softer jets. Figure~\ref{fig:ptmiss_n2x1}~(left) shows the missing transverse momentum distribution obtained with our \PWGPY{} simulation after the cuts listed above are applied. 
%
%%%%%%%%%%%%%%%%%%%%%%%%%%%%%%%%%%%%%%%%%%%%%%%%%%%%%%%%%%
%
%
%%%%%%%%%%%%%%%%%%%%%%%%%%%%%%%%%%%%%%%%%%%%%%%%%%%%%%%%%%
%
Here and in the following, results are presented for the default \NLOPS{} setup obtained by matching the NLO result via the \POWHEG{} formalism with \PYTHIA{}, and for reference also for a LO sample matched with \PYTHIA{} using the same parton-shower settings, referred to as \LOPS{}. The ratios
\beq
R=\frac{d\sigma^\mr{NLO+PS}}{d\sigma^\mr{LO+PS}}\, 
\eeq
help to quantify the impact of the NLO corrections in the presence of parton-shower effects on  distributions of the decay products encountered in the considered reaction. We find that the general features of the NLO corrections are very similar for distributions of the decay particles as for the weakinos produced in the primary hard scattering process, $pp\to \neu_2^0\neu_1^-$. In particular, the $R$ ratio is flat over the entire range of missing transverse momentum, with a size of about 1.2 resembling the ratio of the integrated NLO and LO cross sections.

Figure~\ref{fig:ptmiss_n2x1}~(right) illustrates the invariant mass distribution of the $e^+e^-$ system in the considered process. Apparently, the decay of the $\neu_2^0$ into a lepton pair and the $\neu_1^0$ LSP is dominated by $e^+e^-$ pairs with an invariant mass close to the $Z$ pole. Similarly, the decay of the $\neu_1^+$ into a $\mu^+\nu_\mu$ pair and an LSP features a lepton-neutrino pair dominated by the $W$ resonance. Since the invariant mass of the $\mu^+\nu_\mu$ pair cannot be fully reconstructed because of the non-detectable neutrino we refrain from showing that distribution here. Similar to the case of missing transverse momentum, the $R$ ratio turns out to be flat for the invariant mass distribution of the $e^+e^-$ system, with slightly more statistical fluctuations far away from the resonance region at around $M_{e^+e^-}\sim M_Z$ than in the peak region.

\begin{figure}[t]
\bec
\includegraphics[width=0.49\textwidth]{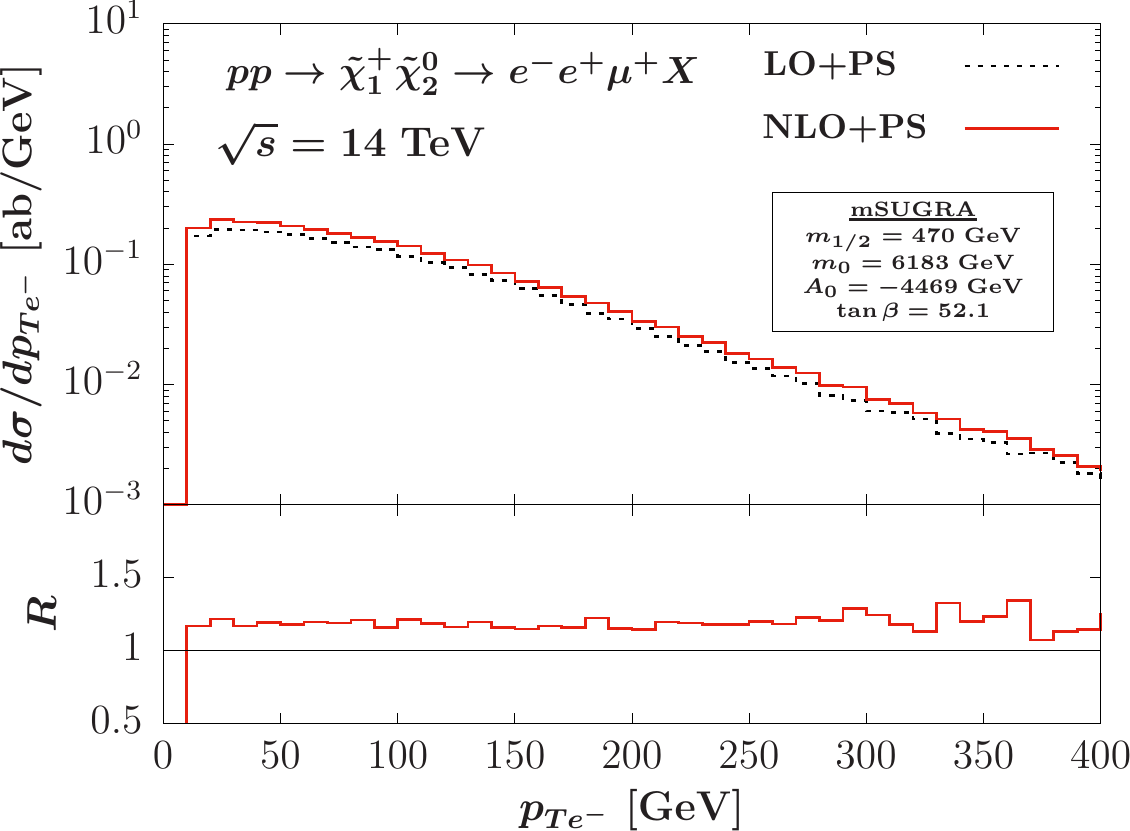}\hspace{2mm}
\includegraphics[width=0.465\textwidth]{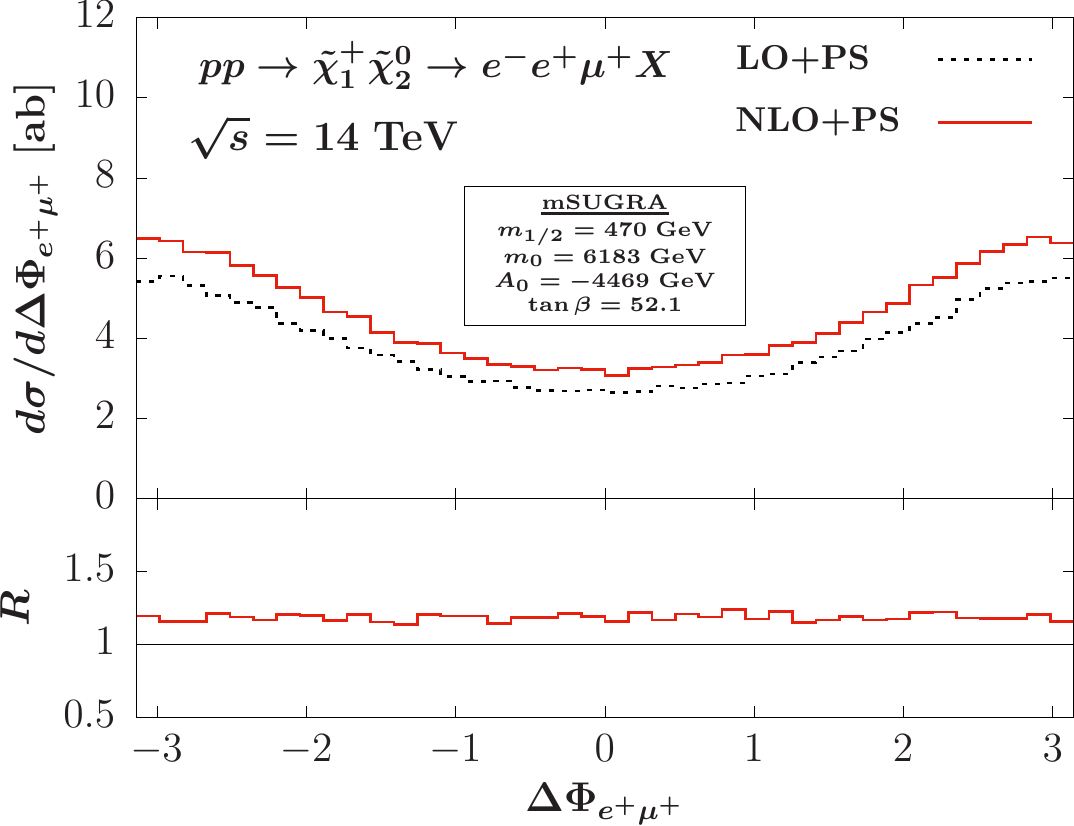}
\caption{
Transverse momentum distribution of the electron (left) and azimuthal-angle separation of the positron and the muon (right) emerging in the $e^+e^-\mu^+ + E_T^\mr{miss}$ decay mode of the process $pp\to \neu_2^0\neu_1^+$ at LO (dotted black lines) and NLO (solid red lines) matched with \PYTHIA{}, after the leptonic cuts of Eqs.~(\ref{eq:lcuts}) and (\ref{eq:ptmiss-cut}) are applied.
}
\label{fig:ptdec_n2x1}
\eec
\end{figure}

The transverse-momentum distribution of the electron is depicted in Fig.~\ref{fig:ptdec_n2x1}. Because of the selection cuts of Eq.~(\ref{eq:lcuts}) that we impose, no events with a transverse momentum smaller than 10~GeV occur. Over the entire plot range, the $R$~ratio amounts to about 1.2,  i.e.\ the NLO corrections  are distributed rather uniformly for this distribution. 
The r.h.s.\ of Fig.~\ref{fig:ptdec_n2x1} shows the azimuthal-angle separation $\Delta\Phi_{e^+\mu^+}$ of the two positively charged leptons occurring in the $e^+e^-\mu^+ + E_T^\mr{miss}$ final state. We note that the azimuthal-angle separation of the positron and the muon peaks at $\pm\pi$. Also for this distribution, the impact of NLO corrections is flat over the entire range considered.
%
%%%%%%%%%%%%%%%%%%%%%%%%%%%%%%%%%%%%%%%%%%%%%%%%%
%
\section{Conclusions}
\label{sec:conc}
In this work, we have presented a new set of implementations for weakino pair-production processes in the framework of the \POWHEGBOX{}. The newly developed code allows for the calculation of the NLO SUSY-QCD corrections for the hard production process, and provides an interface to parton-shower programs such as \PYTHIA{} via the \POWHEG{} method.  The program can process SLHA files obtained with an external spectrum calculator for the computation of a specific SUSY parameter point in the context of the MSSM. If desired, decay chains of the weakinos can be simulated with a dedicated option in  \PYTHIA{}. 

We have described the technical aspects of the implementation specific to weakino pair-production processes. To illustrate the capabilities of the code package we developed, we have discussed phenomenological features of a few selected weakino pair-production processes focusing on theoretical uncertainties and the impact of parton shower effects on experimentally accessible observables. We have found that, in accordance with previous results reported in the literature,  generally NLO corrections have a significant impact on production rates. Scale uncertainties of the NLO results are moderate, however. Parton-shower effects are very small for weakino distributions, but can be significant for jet observables that are effectively described only at lowest order in perturbation theory. Getting better control on these would require a full NLO calculation for weakino pair-production in association with a jet. Finally, we have illustrated the capability of the code to account for the kinematic distributions of observables related to SUSY decay chains for a specific mSUGRA benchmark point. Any user of the code is free, however, to consider a SUSY spectrum of her own choice and obtain \NLOPS{} results for any set of observables within arbitrary selection cuts.

%
%%%%%%%%%%%%%%%%%%%%%%%%%%%%%%%%%%%%%%%%%%%%%%%%%%%%%%%%%%%%%%%
%
\acknowledgments
We are grateful to Tilman Plehn, Michael Spira, and Marco Stratmann for valuable comments and discussions.
This work has been supported in part by the Institutional Strategy of the University of T\"ubingen (DFG, ZUK 63), by the DFG Grant JA 1954/1, and by the German Academic Scholarship Foundation (Studienstiftung des deutschen Volkes). This work was performed on the high-performance computing resources funded by the Ministry of Science,  Research and the Arts  and the Universities of the State of Baden-W\"urttemberg, Germany, within the framework program bwHPC. The Feynman diagrams of this paper have been drawn with the program {\tt JaxoDraw 2.0}~\cite{Binosi:2003yf,Binosi:2008ig}.

%
%%%%%%%%%%%%%%%%%%%%%%%%%%%%%%%%%%%%%%%%%%%%%%%%%%%%%%%%%%%%%%%
%\clearpage

\bibliography{weakino-v2-jhep}{}
\bibliographystyle{JHEP}

\end{document}